\documentclass[12pt]{iopart}
\usepackage{color, graphicx, bm}
\begin{document}

\title{A silicon-based ion trap chip protected from semiconductor charging}

\author{Daun Chung$^{1,2,*}$, Kwangyeul Choi$^{1,2,3,*}$, Woojun Lee$^{1,2,4}$, Chiyoon Kim$^{1,2,3}$, Hosung Shon$^{1,2}$, Jeonghyun Park$^{1,2}$, Beomgeun Cho$^{1,2}$, Kyungmin Lee$^{1,2}$, Suhan Kim$^{1,2,3}$, Seungwoo Yoo$^{1,2,3}$, Eui Hwan Jung$^{1,2,3}$, Changhyun Jung$^{1,2,3}$, Jiyong Kang$^{1,2}$, Kyunghye Kim$^{1,2}$, Roberts Berkis$^{5}$, Tracy Northup$^{5}$, Dong-Il ``Dan'' Cho$^{6}$, Taehyun Kim$^{1,2,3,4,7,8}$}

\address{1. Dept. of Computer Science and Engineering, Seoul National University, Seoul 08826, South Korea}
\address{2. Automation and Systems Research Institute, Seoul National University, Seoul 08826, South Korea}
\address{3. Inter-University Semiconductor Research Center, Seoul National University, Seoul 08826, South Korea}
\address{4. Institute of Computer Technology, Seoul National University, Seoul 08826, South Korea}
\address{5. Dept. of Experimental Physics, University of Innsbruck, Technikerstra\ss{}e 25, 6020 Innsbruck, Austria}
\address{6. Dept. of Electrical and Computer Engineering, Seoul National University, Seoul 08826, South Korea}
\address{7. Institute of Applied Physics, Seoul National University, Seoul 08826, South Korea}
\address{8. NextQuantum, Seoul National University, Seoul 08826, South Korea}
\address{* These authors contributed equally to this work.}
\ead{taehyun@snu.ac.kr}
%\vspace{10pt}
%\begin{indented}
%\item[]September 2024
%\end{indented}

\begin{abstract}
Silicon-based ion trap chips can benefit from existing advanced fabrication technologies, such as multi-metal layer techniques for two-dimensional architectures and silicon photonics for the integration of on-chip optical components. However, the scalability of these technologies may be compromised by semiconductor charging, where photogenerated charge carriers produce electric potentials that disrupt ion motion. Inspired by recent studies on charge distribution mechanisms in semiconductors, we developed a silicon-based chip with gold coated on all exposed silicon surfaces. This modification significantly stabilized ion motion compared to a chip without such metallic shielding, a result that underscores the detrimental effects of exposed silicon. With the mitigation of background silicon-induced fields to negligible levels, quantum operations such as sideband cooling and two-ion entangling gates, which were previously infeasible with the unshielded chip, can now be implemented.
\end{abstract}

\section{Introduction}
\label{sec_intro}

\begin{figure*}[ht]
\centering
\includegraphics[width=1\textwidth]{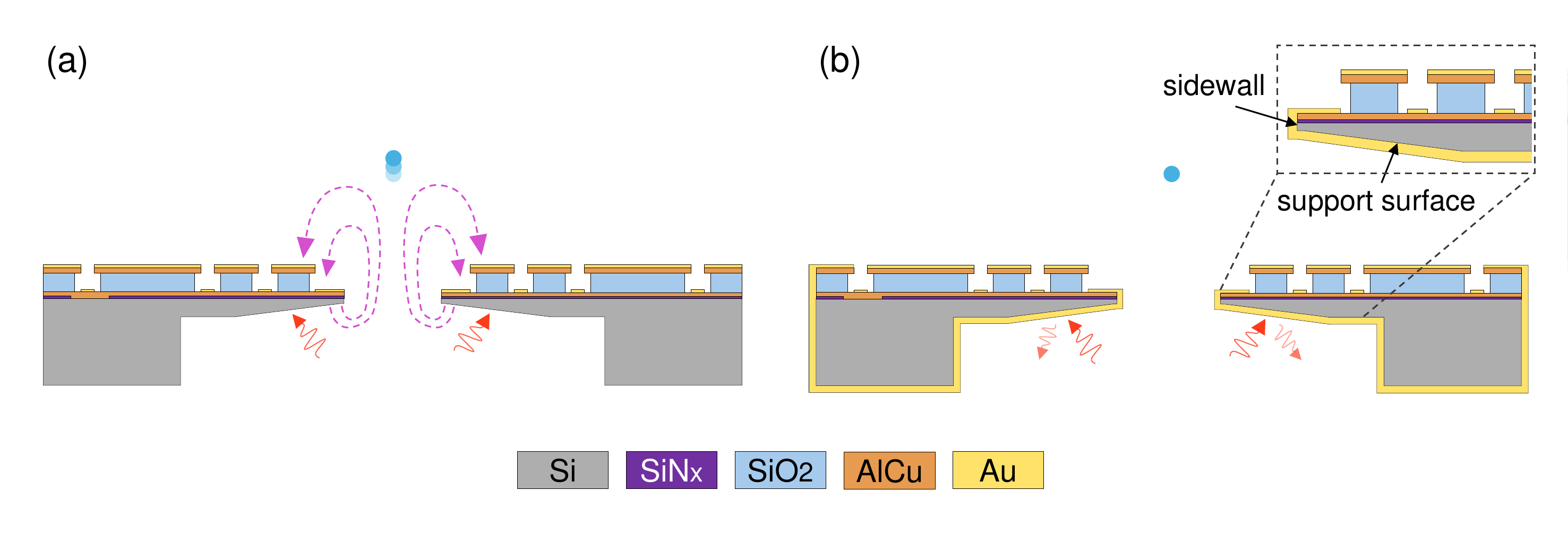}
\caption{Cross-sectional view of the ion trap chips (a) with and (b) without gold coating. A protective gold layer is added on all exposed silicon surfaces in the new chip. The thickness is exaggerated for visual aid. The wavy arrows indicate light impinging on the surfaces of the slot structure. In (a), silicon-induced stray fields are represented as purple dashed arrows. The surfaces of the slot structure consist of the sidewall and support surface as shown in the inset of (b).}
\label{fig_intro_chip_cross-section}
\end{figure*}

Ion trap systems are among the leading platforms for quantum computation, as demonstrated by achievements in high-fidelity operations~\cite{debnath_2016, shaffer_2018, wright_2019, postler_2022, mayer_2024, loschnauer_2024, chen_2024}. However, scalability remains a critical challenge~\cite{monroe_2013, guo_2024}, requiring thorough consideration of numerous factors, including even the material of the trap~\cite{brown_2021}. As efforts to develop optimal ion trap architectures continue, silicon-based surface traps have been demonstrated as a strong candidate for achieving such scalability~\cite{Stick_2006, clark_2021, Blain_2021, moses_2023}. 

Their advantage is leveraged by mature semiconductor fabrication technologies, which allow for flexible trap design and integration of on-chip components essential for scalability. For instance, arbitrary chip shapes and junction fabrication are enabled by multi-metal layer structures where complex electric traces are routed in a three-dimensional layout beneath the chip surface \cite{Blain_2021}. This also facilitates ion shuttling, a key technique for realizing the quantum charge-coupled device (QCCD) architecture proposed for large-scale operations~\cite{kielpinski_2002, pino_2021}. Additionally, direct integration of optical and electrical components for on-chip beam delivery and detection is actively being pursued with ion trap fabrication \cite{Mehta_2020, Niffenegger_2020, Kwon_2024}. 

While the benefits provided by this technological foundation are expected to outweigh certain drawbacks of surface traps compared to macroscopic traps---such as lower trap depth~\cite{Chiaverini_2005}, increased susceptibility to dielectric charging~\cite{Harlander_2010}, higher heating rates~\cite{Sedlacek_2018_2}, and greater anharmonicity~\cite{Home_2011}---there is a significant yet poorly understood issue specific to semiconductor-based surface traps. This issue, known as semiconductor charging,
involves the photogeneration of charge carriers and their subsequent dynamics throughout the bulk and surfaces of a semiconductor~\cite{Lee_Chung_2024}. Such charge carriers can generate strong stray electric fields at the ion position, scrambling the quantum evolution of motion-sensitive operations. 

Recently, a microscopic model for the charge distribution mechanism and generation of surface photovoltage has been presented and experimentally validated in a chip with exposed silicon surfaces~\cite{Lee_Chung_2024}. Optical shielding of the exposed surfaces was suggested as a potential method to mitigate semiconductor charging by suppressing photogenerated charge carriers, a chip with this feature was not previously demonstrated and tested to confirm the conjecture. In this work, we present results obtained from a chip specifically fabricated to verify this mitigation method and assess its effectiveness.

The new fabricated chip features metallic shielding on all exposed silicon surfaces as shown in figure~\ref{fig_intro_chip_cross-section} (b). Considerable effort is focused on coating the surfaces at the slot structure---the sidewall and the support surface---as these areas are sources of substantial surface photovoltage (SPV) that originates from interface states formed during the deep reactive-ion etching (DRIE) process \cite{Lee_Chung_2024, Jung_2021, mohammed_surf_2005}. While the support surface is relatively easy to coat uniformly with gold, it is challenging to achieve full coverage on the sidewalls due to scallop patterns created during the DRIE process. A detailed description of the techniques associated with the removal of scallop patterns is provided in section~\ref{sec_fabrication}. Although it has been quite common to shield non-metallic surfaces with gold in ion trap chips, including silicon chips, to improve dielectric shielding and reduce ion heating \cite{Blain_2021, Mount_2013}, our approach is distinguished by its focus on suppressing semiconductor charging. 
% Since the DRIE process is a widely used technique for shaping and cutting through silicon wafers, we anticipate that our solution will have broad applicability.

The paper is structured as follows: Section~\ref{sec_fabrication} outlines the overall fabrication process of the ion trap chip. Section~\ref{sec_field_measurement} demonstrates the effectiveness of our methods through experimental data that show suppressed silicon-induced stray fields and stabilized ion motion. Finally, section~\ref{sec_operation} presents results of quantum operations, including sideband cooling~\cite{schmidt-kaler_2002}, heating rate measurements~\cite{Diedrich_1989}, and the implementation of the M{\o}lmer-S{\o}rensen gate on a pair of $^{171}$Yb$^+$ ions~\cite{Sorensen_1999}, all of which were not feasible in our previous chip without metallic shielding~\cite{Lee_Chung_2024}.

\section{Trap fabrication}
\label{sec_fabrication}

\begin{figure*}[t]
\centering
\includegraphics[width=1\textwidth]{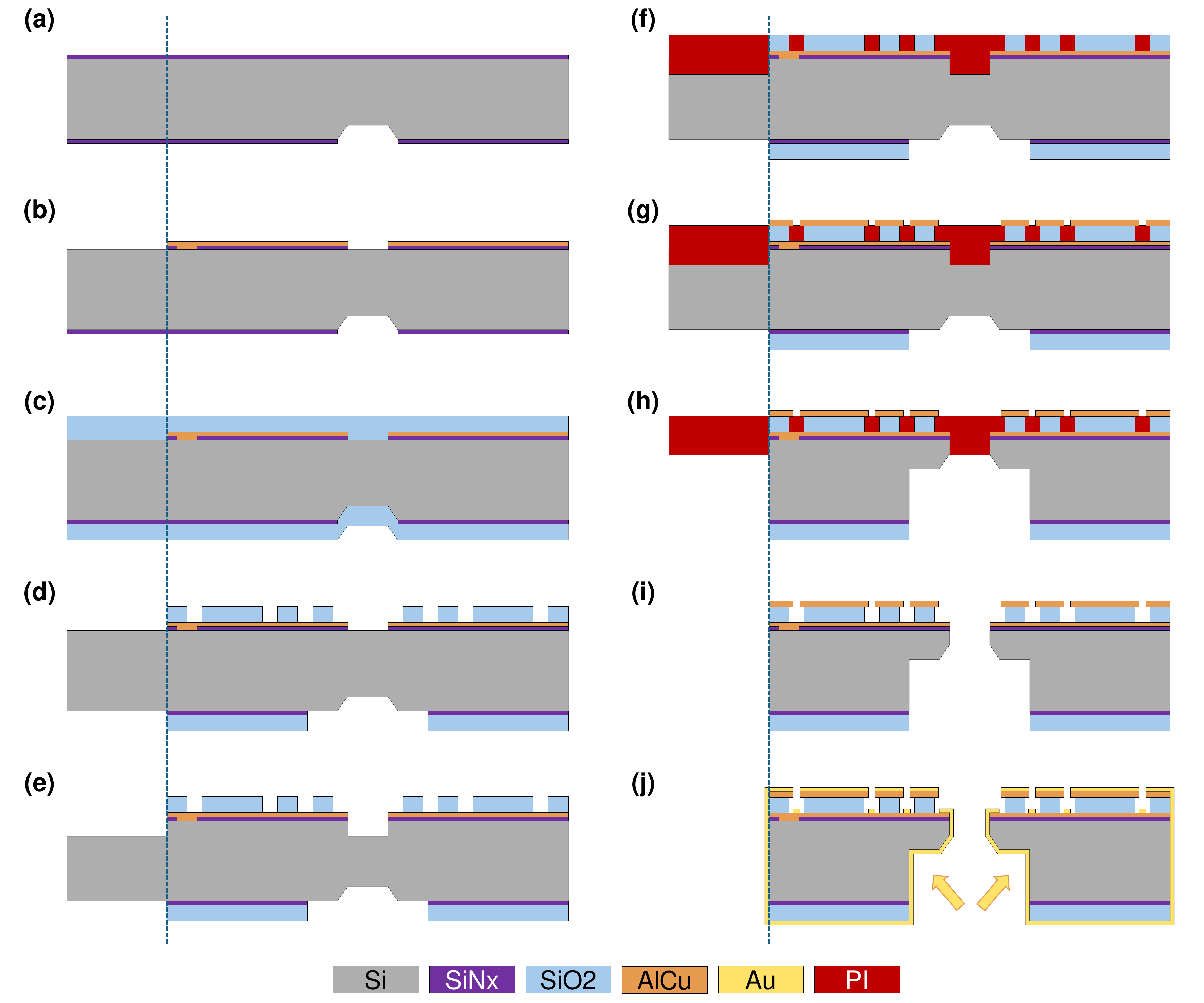}
\caption{Fabrication process flow of the ion trap chip. It also illustrates the formation of the complex chip outline, where the vertical dashed lines represent the boundary of the chip. To the left of these lines, the area is etched to create the complex outer boundary, designed to minimize laser scattering. To the right of the vertical lines, a cross-sectional view of the chip shows the internal layers and structure formed during the fabrication process (not drawn to scale).}
\label{fig_process_chart}
\end{figure*}

% This chapter outlines the key elements of the fabrication process for microstructures designed for ion trapping. The focus is on three major components: the detailed process flow, the use of scallop smoothing techniques, and the advantages of using SOI wafers. The fabrication process closely follows the methodology outlined in reference~\cite{Jung_2021}, with only slight modifications. The core steps, such as SiN$_\mathrm{x}$ deposition, KOH wet etching, and DRIE etching, remain largely the same. However, a notable addition in this process is the incorporation of a gold (Au) deposition step, which enhances the electrical conductivity and durability of the ion-trapping electrodes. Additionally, an extra process was applied to ensure complete gold coverage, particularly addressing challenges like uneven coating on scalloped surfaces, further improving the performance and reliability of the final device compared to earlier methods.

This section elaborates on the new fabrication process of the ion trap chip. The fabrication process involves a series of dielectric deposition, etching, and metal layer formation to produce the desired microstructures. The process flow is presented in figure~\ref{fig_process_chart}.

The process begins with an anisotropic wet etch using potassium hydroxide (KOH) on the backside of the silicon wafer (a). The SiN$_\mathrm{x}$ layer serves as a protective mask during this step, allowing for anisotropic etching to define angled sidewalls on the wafer’s backside. The primary function of the KOH etch at this stage is to shape angled sidewalls on the exposed silicon, which later defines the geometry of the slot structure. Following the KOH etching, an aluminum-copper (AlCu) layer is deposited and patterned (b). This metal layer acts as a ground and shielding layer between the silicon substrate and the metal layers to be deposited in step (g). 

A silicon dioxide (SiO$_\mathrm{2}$) layer of 10 µm thickness is deposited next (c), acting as an insulating barrier between the conductive AlCu layers and other structures. This layer is then patterned using dry etching (d).  mploys DRIE on the front side of the wafer to control the remaining silicon thickness to around 10 µm. In step (f), polyimide is coated and subsequently planarized using chemical mechanical planarization (CMP). The polyimide serves as a sacrificial layer for the subsequent steps. In step (g), AlCu is again deposited and patterned to form the electrodes, intentionally extending beyond the underlying SiO$_\mathrm{2}$ pillars to create an overhang structure. While charges can accumulate on the SiO$_\mathrm{2}$ surface, the AlCu overhang is designed to shield these charges from the trapped ions~\cite{Stick_2006}. 

In step (h), DRIE is performed on the backside of the wafer to form a slot structure, which provides additional laser pathways. This step also completely etches the outer boundary of the chip, forming a complex-shaped outline that can minimize the scattering of lasers propagating parallel to the chip's surface. Further details on the formation of the chip’s outer boundary are discussed in section~\ref{sec_outline}. In step (i), the polyimide layer is stripped away, exposing the underlying structures, including the silicon within the slot structure. Once the polyimide is removed, the exposed silicon surfaces on the sidewalls of the slot structure undergo a scallop smoothing process to improve surface quality. The details of this scallop smoothing process are elaborated on in section~\ref{sec_scallop}.

The deposition of the gold (Au) layer is carried out in two stages (j), targeting both the front and back sides of the wafer. During the gold evaporation for the backside deposition, the wafer is tilted to ensure that the gold fully covers the exposed silicon within the slot structure. The gold layer on the top prevents oxidation of the underlying aluminum layer, while the gold on the bottom ensures comprehensive coverage of the exposed silicon.

\subsection{Scallop smoothing}
\label{sec_scallop}

\begin{figure*}[t]
\centering
\includegraphics[width=1\textwidth]{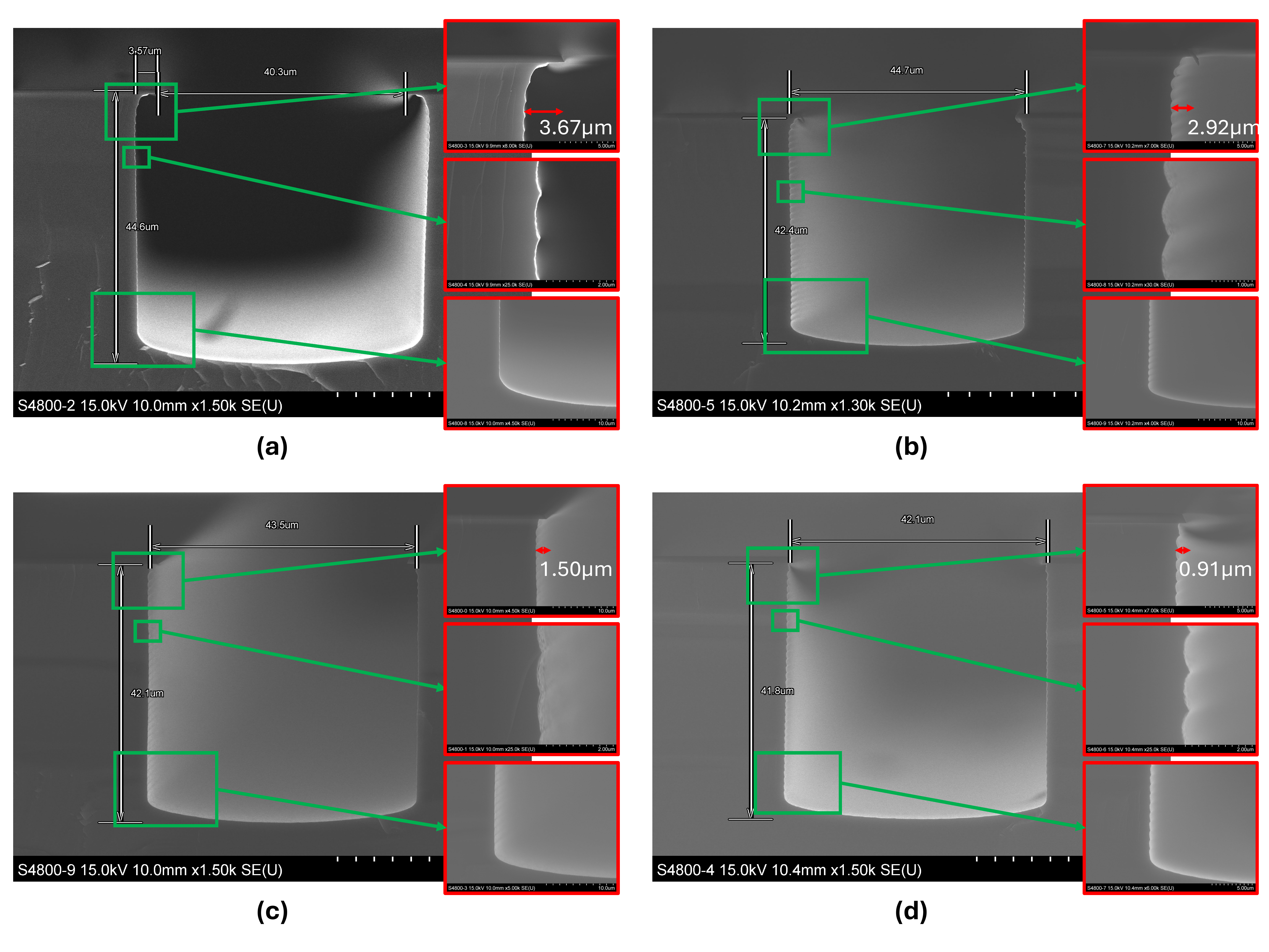}
\caption{Scanning electron microscope (SEM) images of silicon sidewalls after scallop smoothing using different RIE process conditions. (a): RF power: 200 W, SF$_\mathrm{6}$: 16 sccm, O$_\mathrm{2}$: 8 sccm, pressure: 200 mTorr, 600 seconds. (b): RF power: 200 W, SF$_\mathrm{6}$: 16 sccm, O$_\mathrm{2}$: 8 sccm, pressure: 200 mTorr, 300 seconds. (c): RF power: 200 W, SF$_\mathrm{6}$: 30 sccm, pressure: 200 mTorr, 300 seconds. (d): RF power: 100 W, SF$_\mathrm{6}$: 30 sccm, pressure: 200 mTorr, 300 seconds.}
\label{fig_scallop_smoothing_test}
\end{figure*}

\begin{figure*}[t]
\centering
\includegraphics[width=1\textwidth]{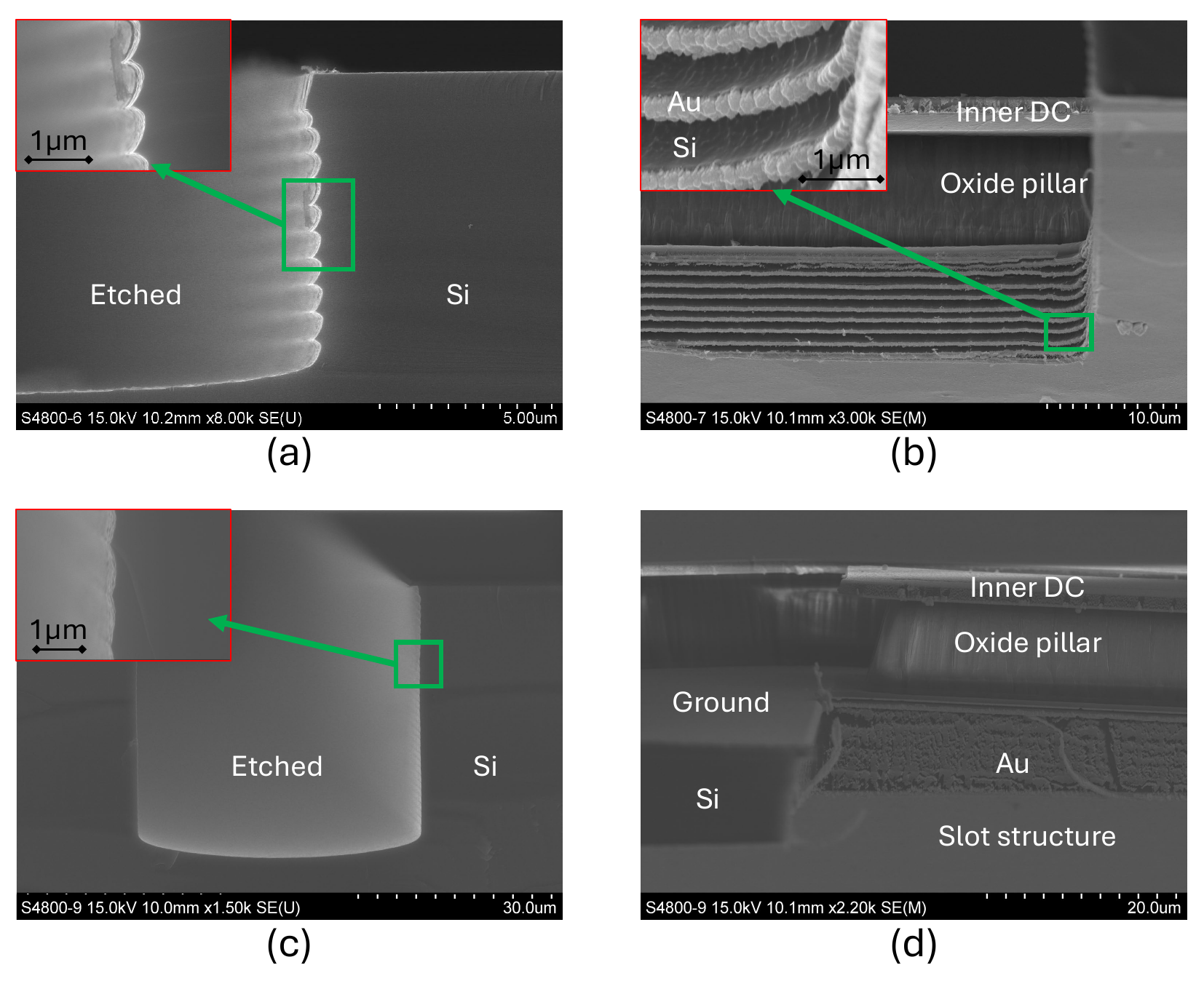}
\caption{Comparison of scallop smoothing and gold deposition at slot structure on ion trap chip using Si wafers. In (a), the image highlights the scalloped sidewall without smoothing. In (b), the image depicts a chip without scallop smoothing after gold deposition, where incomplete gold coverage is evident due to the shadow effect. In contrast, (c) shows the sidewall after scallop smoothing, showing a much smoother surface. (d) illustrates the chip with scallop smoothing after gold deposition, where the gold has been relatively evenly and completely deposited across the smoothed sidewalls.}
%20240402 폴더에 있는 사진 참조, (a): grrcrecipe, (b): test1, (c): test2, (d): test3
\label{fig_scallop_smoothing}
\end{figure*}

During the DRIE process, scallop patterns commonly emerge on the sidewalls of etched silicon due to the alternating etch and passivation cycles. While these scallop patterns are a typical characteristic of DRIE, they pose significant challenges during the subsequent gold deposition step. The scallop-shaped surface can prevent complete coverage of the silicon sidewalls formed during step (e) in figure~\ref{fig_process_chart}, as the irregularities cause a shadowing effect, leaving certain areas of silicon exposed.

To address this issue, an additional process is introduced to smooth the scallop patterns on the silicon sidewalls before gold deposition~\cite{Park2020}. This step involves an isotropic etching process, which etches away silicon uniformly in all directions. By applying this isotropic etch, the sharp peaks and valleys of the scallop patterns are reduced, creating a smoother surface. 

% The smoothing effect ensures the full coverage of gold on the silicon sidewalls, eliminating the discontinuity of the gold coating that would otherwise occur due to the scallop pattern.

The scallop smoothing process is tested under four different reactive-ion etching (RIE) conditions using Oxford Instruments PlasmaPro 80 RIE, as shown in figure~\ref{fig_scallop_smoothing_test}. To establish optimal process conditions, we first use a DRIE process to etch a standard silicon wafer to a depth of 40 µm with a pattern size of 40 µm, employing a 1-µm SiO$_\mathrm{2}$ layer as a mask. Following this, then RIE processes described in the caption of figure~\ref{fig_scallop_smoothing_test} are applied. Notably, condition (c) yields the best results; the isotropic etching applied in this condition effectively removes the scallop patterns and minimizes undercutting--- critical for maintaining the integrity of the slot structure. This condition produces the smoothest sidewalls with minimal undercutting (1.5 µm), ensuring that the subsequent gold deposition achieves full coverage across the silicon sidewalls.

%The etching is performed on a standard silicon wafer, with a 1--µm SiO$_\mathrm{2}$ layer used as a mask for the DRIE process. The pattern size was 40 µm, with an etch depth of 40 µm. After the DRIE etch, the scallop smoothing process is applied. While detailed conditions are provided in the figure caption, it is worth noting that (c) yield the best results. The isotropic etching applied in (c) not only removes the scallop patterns effectively but also minimized undercutting, which is critical for maintaining the integrity of the slot structure. This condition produces the smoothest sidewalls with the least amount of undercut(1.5 µm), ensuring that the gold deposition achieved full and uniform coverage across the silicon sidewalls.

In figure~\ref{fig_scallop_smoothing} (a) and (b), the challenges associated with the DRIE-induced scallop patterns on the silicon's sidewalls are illustrated. Figure~\ref{fig_scallop_smoothing} (a) highlights the characteristic scallop patterns that arise from the DRIE process, leading to uneven surface topography. 
%This irregular surface significantly complicates subsequent gold deposition, as the scallop patterns create shadowing effects, preventing full coverage. 
Figure~\ref{fig_scallop_smoothing} (b) shows the result of gold deposition on an unsmoothed surface. The presence of scallop patterns results in incomplete gold coverage, with thinner or entirely uncoated regions. This incomplete gold coating may lead to the formation of photogenerated charge carriers at the slot structure.

In contrast, figure~\ref{fig_scallop_smoothing} (c) illustrates the silicon sidewalls after isotropic etching, which smooths the scallop patterns and reduces surface irregularities. This results in a more uniform surface that facilitates consistent gold deposition, as shown in figure~\ref{fig_scallop_smoothing} (d).

Although achieving full gold coverage on the sidewalls of the slot structure in the chip was challenging, the support surface (see figure~\ref{fig_intro_chip_cross-section} (b)) was uniformly coated with gold, shielding the largest portion of the exposed silicon responsible for semiconductor charging~\cite{Lee_Chung_2024}. The overall fabrication process described in this section proved to be highly effective, as demonstrated in sections~\ref{sec_field_measurement} and~\ref{sec_operation}.

\subsection{Fabrication of a chip with an arbitrary outline}
\label{sec_outline}

\begin{figure*}[t]
\centering
\includegraphics[width=1\textwidth]{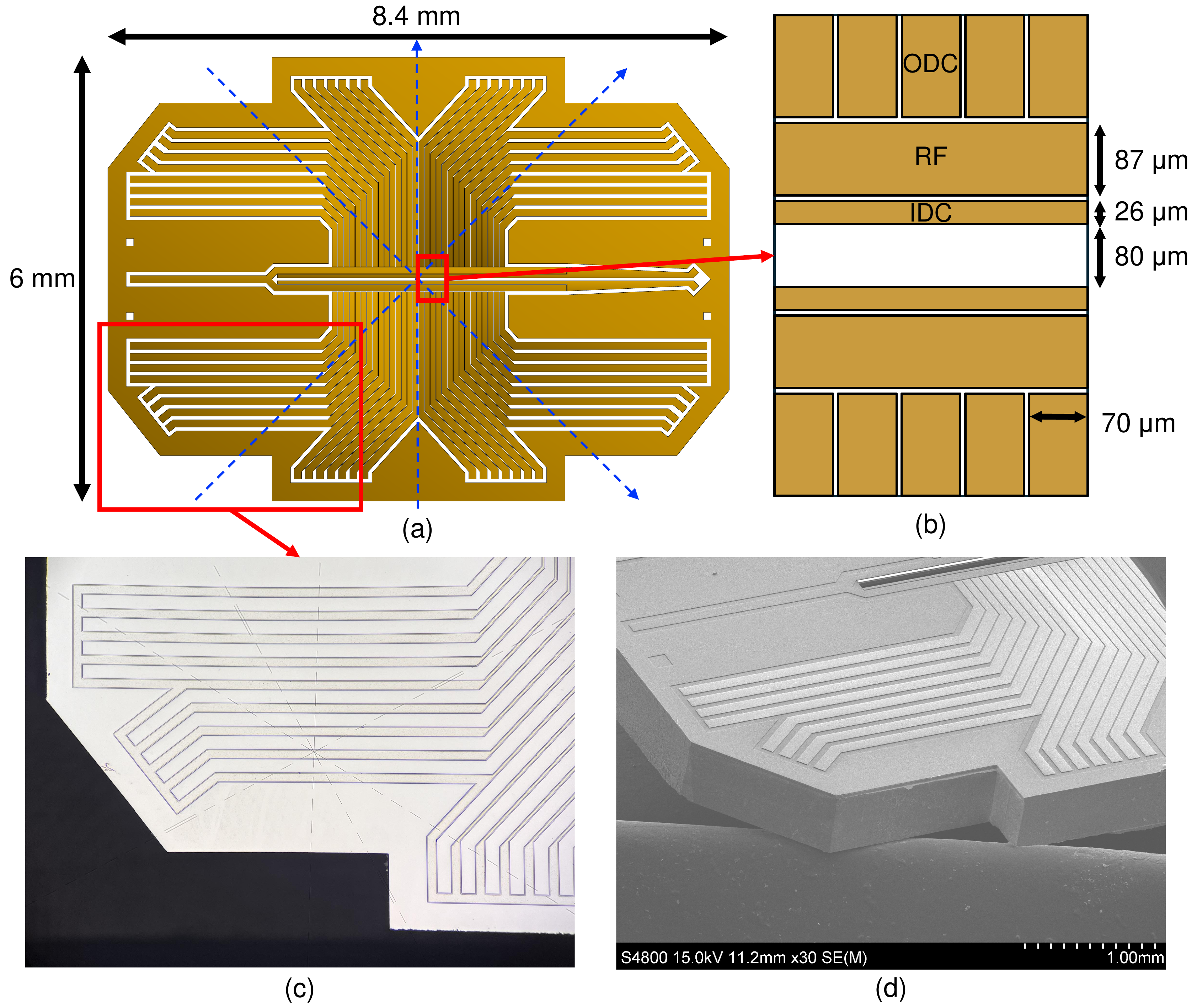}
\caption{Images of the complex chip outline designed to minimize laser clipping. In (a), a schematic of the chip electrode layout is shown. Blue dashed lines represent the potential beam paths. The electrode dimensions are listed in (b). Electrodes are separated by gaps of width 8 µm. In (c), an optical microscope image shows the complex outer boundary of the chip, clearly displaying the design aimed at reducing laser clipping and scattering. The dashed lines in this image are artifacts of the optical microscope marking the origin and hold no significance. In (d), an SEM image taken at a similar magnification shows the outer boundary from a tilted perspective, revealing the cleanly etched edges of the outline. }
\label{fig_arbitrary_outline}
\end{figure*}

In ion trap systems, precise control of the laser position is crucial for stable trapping and quantum gate implementation. One of the challenges arises when lasers, mostly aligned parallel to the chip surface and aimed at ions positioned at heights of typically 40–150 µm, cause scattering and clipping~\cite{Hong_2016}. Long laser paths across the chip surface increase the likelihood of laser interaction with chip features, leading to scattering of the laser.

To address this issue, we have developed a new process for fabricating a chip with complex, arbitrary outlines designed to minimize clipping and reduce laser scattering. While chips with complex outlines have previously been reported~\cite{Blain_2021, moses_2023}, we are not aware of any publications that detail their specific fabrication process. 
% By employing specific process techniques, we shape the outer boundary of the chip to reduce laser scattering from the chip surface. 
The design of the new chip is exhibited in figure~\ref{fig_arbitrary_outline} (a). The beam paths are represented by blue dashed lines. The dimensions of the electrodes responsible for generating trapping potentials are denoted in (b), where IDC, RF, and ODC stand for inner-DC, radiofrequency, and outer-DC electrodes, respectively~\cite{house_2008}. The complex chip outline is formed by selectively etching the regions to the left of the vertical green dashed lines shown in figure~\ref{fig_process_chart}.

% This etching process precisely removes material to create the desired complex outer shape.  

Importantly, while most of the outer boundary is defined during steps (d) to (h) in figure~\ref{fig_process_chart}, complete singulation of the chip does not occur until the final dicing step. Conducting full outline etching using DRIE at step (h) would lead to premature singulation, disrupting subsequent wafer-scale fabrication steps and necessitating processing on individual dies. However, by etching only part of the outline at this stage, the remaining fabrication processes can continue at the wafer level. This approach preserves the integrity of the wafer until the final dicing step, ensuring that singulation occurs only after most of the fabrication processes are completed.

In figure~\ref{fig_arbitrary_outline}, the results of the etched chip outline are shown. Figure~\ref{fig_arbitrary_outline} (c) presents an optical microscope image from a top-down view, displaying the clean-cut outer boundary. Figure~\ref{fig_arbitrary_outline} (d) is an SEM image taken at an angle, highlighting the etched outlines achieved through the fabrication process. Notably, no residual structures or unwanted protrusions were formed during the etching, and we obtain a clean and sharp outer boundary. These images demonstrate the effectiveness of the fabrication process in creating the precise and detailed outline required to minimize laser clipping.

\section{Silicon-induced stray field measurement}
\label{sec_field_measurement}
\begin{figure*}[ht]
\centering
\includegraphics[width=1\textwidth]{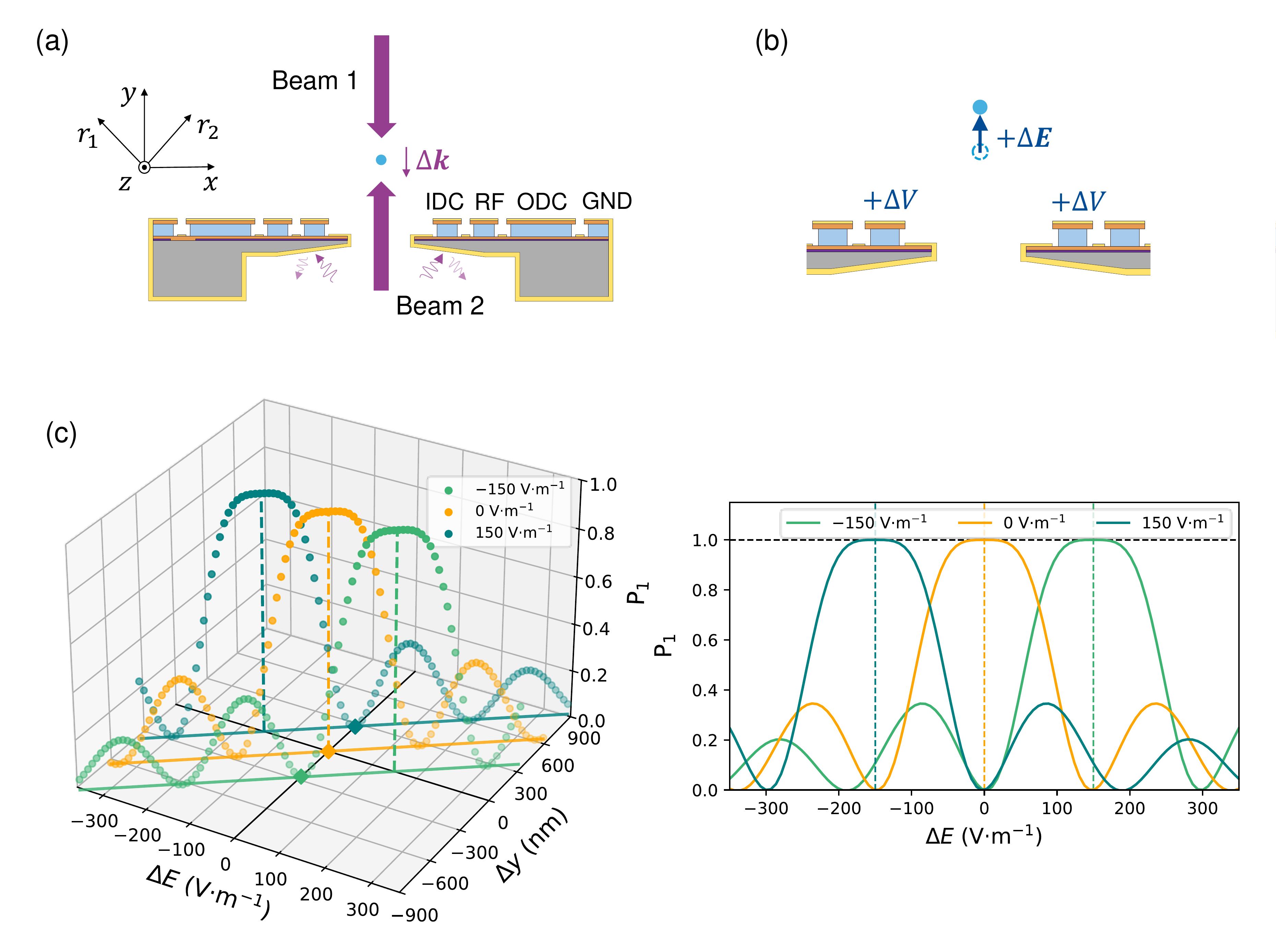}
\caption{Experiment scheme for measuring silicon-induced stray fields. (a) Counter-propagating Raman beam configuration. IDC, RF, ODC, and GND stand for inner DC, radiofrequency, outer DC, and ground electrodes, respectively (see section~\ref{sec_outline}). (b) Schematic of the DC scanning method. A voltage $\Delta V$ is applied to the IDC electrodes and scanned. (c) Visualization of the stray electric field measurement via the DC scanning method. (Left) A three-dimensional plot about the compensation field $\Delta E$, the ion displacement $\Delta y$, and the qubit transition probability $\mathrm{P}_1$, given some background stray field $E_{y}$ denoted in the legend. (Right) The projection of the three-dimensional plot onto the $\Delta E$-$\mathrm{P}_1$ plane. Bessel-type patterns occur due phase modulation arising from excess micromotion. The maximum of $\mathrm{P}_1$ occurs when the compensation field exactly cancels out the stray field.}
\label{fig_experiment_scheme}
\end{figure*}

\begin{figure*}[ht]
\centering
\includegraphics[width=1\textwidth]{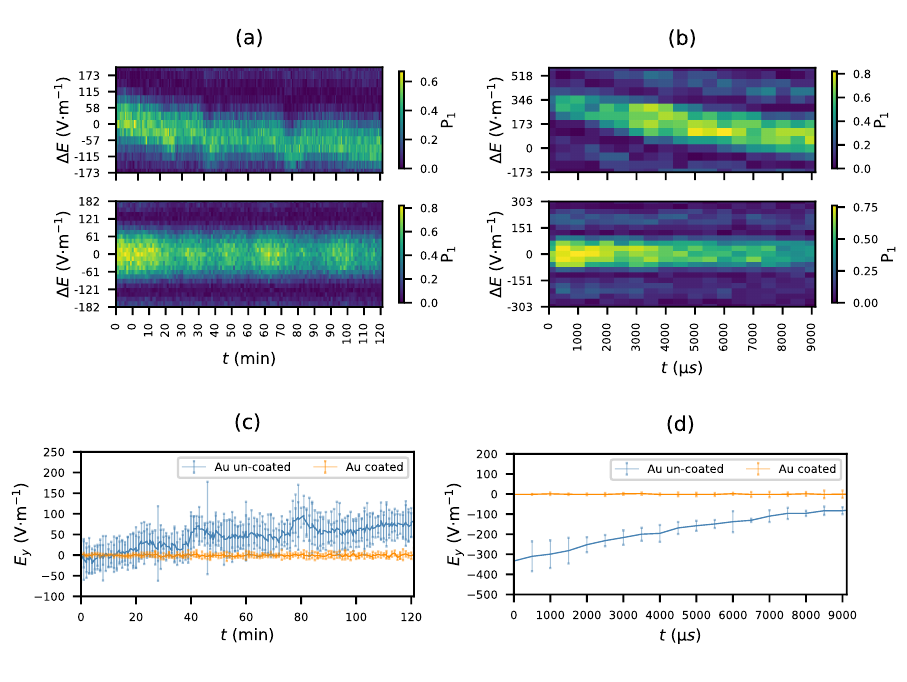}
\caption{Stray field measurement results from ion trap chips with and without metallic shielding. (a), (c) Long term monitoring and (b), (d) short term evolution of the stray fields. The bottom and top plots in (a) and (b) correspond to the chip with and without gold coating, respectively. $E_y$ represents the stray field along the direction normal to the chip surface. In (c) and (d), the compensation field values extracted from fitting the data in (a) and (b) to Bessel-type profiles are overlaid to directly compare the stray fields between the two chips.}
\label{fig_stray_field_measurement}
\end{figure*}

In the chip with no metallic shielding, a peculiar optical absorption spectrum attributed to the physical properties of the defect states at the silicon surface has been reported~\cite{Lee_Chung_2024}. There is strong absorption in the near-infrared range near the silicon band gap, with a monotonically weaker response extending through the visible and ultraviolet ranges. Photon flux in the range of \( 10^{14} - 10^{16} \ \mathrm{cm}^{-2} \cdot \mathrm{s}^{-1} \) is sufficient to induce significant stray fields, depending on the wavelength. In the new chip with a protective gold layer, however, no such charging is observed across all previously tested wavelengths, even at higher power levels. This observation serves as evidence that charge carrier dynamics at the exposed silicon surface are responsible for the photoinduced charging observed in the chip without gold coating, further solidifying the experimental verification of semiconductor charging.

In order to characterize the reduced semiconductor charging more quantitatively, we use the DC scanning method to directly measure the silicon-induced stray fields~\cite{Lee_2023}. The experimental scheme is shown in figure~\ref{fig_experiment_scheme}. We use Doppler-cooled $^{171}$Yb$^+$ ions and a mode-locked laser with a center wavelength of 355 nm to drive a Raman transition~\cite{Mizrahi_2014}. As depicted in figure~\ref{fig_experiment_scheme} (a), two counter-propagating beams with a wave vector difference $\Delta \textbf{\textit{k}}$ oriented perpendicular to the chip are delivered to the ion through the slot structure. This configuration makes the Raman transition sensitive to electric fields along the $y$-axis.  For the chip without gold coating, as shown in figure~\ref{fig_intro_chip_cross-section} (a), semiconductor charging is primarily caused by beam 2, since the silicon surface can see its scattered light directly (see figure~\ref{fig_experiment_scheme} (a)). Note that $\Delta \textbf{\textit{k}}$ overlaps with both radial modes ($r_{1},\ r_{2}$) of the ion. 

The probe electrodes in the DC scanning method are the IDC electrodes shown in figure~\ref{fig_experiment_scheme} (a). As can be seen in figure~\ref{fig_experiment_scheme} (b), a voltage $\Delta V$ is applied to the IDC electrodes and scanned. This generates an electric field $\Delta \bm{E}$, defined as the compensation field, at the ion position. When $\Delta V$ is applied symmetrically to both IDC electrodes, the resulting compensation field, denoted as $\Delta E$, is produced only along the $y$-axis. A visual description of the method is presented in figure~\ref{fig_experiment_scheme} (c), where numerical simulations have been performed with parameters detailed in section~\ref{sec_operation}. In the three-dimensional plot, $\Delta y$ is the displacement of the ion along the $y$-axis with respect to the RF null, and $\mathrm{P}_1$ is the qubit transition probability. Three cases are shown, each with a different background stray field value $E_{y}$, as denoted in the legend. The markers along the $\Delta y$-axis represent the ion displacement due to $E_{y}$. 

Away from the RF null, the ion undergoes excess micromotion, which induces variations in the transition strength of the Raman transition. When the carrier transition is driven for a $\pi$-pulse duration as the compensation field $\Delta E$ is varied, the excitation probability $\mathrm{P}_1$ is modified, resulting in a Bessel-type pattern~\cite{Lee_2023}. The maximum occurs when the compensation field exactly cancels out the stray field, $\Delta E = -E_y$, restoring the ion back to the RF null. Note that the vibrational states of the ion have been assumed to be in the ground state for simplicity. Even when there are vibrational states other than the ground state, the maximum occurs at the same value of $\Delta E$~\cite{Lee_2023, Lee_Chung_2024}. The plot on the right in figure~\ref{fig_experiment_scheme} (c) shows the projection of the three-dimensional plot onto the $\Delta E$-$\mathrm{P}_1$ plane, which corresponds to what is measured in real experiments. The stray field $E_{y}$ can thus be monitored by tracking the offset of the Bessel-type pattern as the compensation field $\Delta E$ is scanned.

Both long-term field drift and fluctuations, as well as short-term field generation when the laser is switched on ~\cite{Lee_Chung_2024, Lee_2023}, are measured by applying the DC scanning method over time. The results are shown in figure~\ref{fig_stray_field_measurement}, where (a), (c) correspond to the long-term monitoring and (b), (d) the short-term response immediately after a laser is turned on. The bottom and top figures in (a) and (b) are obtained from the chip with and without gold coating, respectively. By fitting each column of the time series data to a Bessel function with an offset (as shown in Figure~\ref{fig_experiment_scheme} (c)), we determine the value of the stray field and its uncertainty from the fitted offset value, $\Delta E_\mathrm{fit} = - E_{y}$. The fitted values extracted from (a) and (b) are plotted in (c) and (d), respectively, where the error bars represent the uncertainty of the fit at each instance. The magnitude of the stray field is calculated with the COMSOL Multiphysics\textsuperscript{\tiny\textregistered} software. 

As shown in figure~\ref{fig_stray_field_measurement} (a) and (c), the long-term drift and fluctuations are suppressed to a negligible level. The error bars are also considerably smaller than those in data from the unshielded chip, where rapid stray field fluctuations occurring within the duration of a single DC scan increase the uncertainty of the fit. In the absence of such fast varying fields in the shielded chip, we are able to estimate the stray field with greater precision. The small fluctuations over time in (c) can be explained by drifts in beam parameters, such as beam pointing, which alters the $\pi$-pulse duration at each instance. These fluctuations also explain the variations of the transition strength over time in (a). The short-term field build up is also greatly suppressed, as shown in (b) and (d). Experiments with the chip lacking metallic shielding typically requires a pre-turn on procedure~\cite{Lee_Chung_2024}, where beam 2 is turned on for the field generation time prior to turning on beam 1, so that the Raman transition is driven after the ion has reached a new equilibrium position. As shown in (d), this sequence is unnecessary in the new chip as there is no observable photoinduced stray field.

% Note that semiconductor charging is effectively mitigated even with the incomplete protective gold layer shown in figure~\ref{fig_scallop_smoothing} (d). We believe that switching to a more highly conductive silicon wafer (p-type doping concentration of $10^{15} \ \mathrm{cm}^{-3}$ to $10^{19} \ \mathrm{cm}^{-3}$) has also helped by lowering the Fermi level, hence reducing the number of optically excitable defect states at the surface~\cite{Lee_Chung_2024}. The partially exposed silicon surfaces may still, however, contribute to the high heating rate measured in our system (see section~\ref{sec_operation}), as electric field noise from space charge regions~\cite{Ghibaudo_1987, Izpura_2007}, generation--recombination processes~\cite{Bonani_1999}, and Schottky barriers~\cite{Hsu_1971} are well known. With our latest developments in gold coating fabrication that achieves nearly full coverage we expect improvements in the ion heating as well. 

% Some remarks can be made on the notable differences in transition strengths over time between the two chips in figure~\ref{fig_stray_field_measurement} (c). In the previous chip, the excitation probability increased as the stray field approaches zero, whereas in the new chip, it gradually decreased over time. This is due to the Doppler cooling beam, which was turned on during the pre-turn on time in the former, and turned off in the latter. Therefore, the decay in transition strength observed in the new chip is due to heating.

\section{Quantum operations as benchmarks of chip performance}
\label{sec_operation}
The elimination of photoinduced stray fields leads to stable ion motion, enabling precise control of motion-sensitive operations. First, phase modulation of the Rabi frequency due to excess micromotion is removed~\cite{Lee_2023}. Also, secular frequency shifts associated with stray field fluctuations, which are problematic in the unshielded chip, are not observed in the shielded chip. Such improvements allow the characterization of vibrational modes, sideband cooling, and qubit control via M{\o}lmer-S{\o}rensen interactions. 

\subsection{Sideband cooling and heating rate measurement}
\begin{figure*}[ht]
\centering
\includegraphics[width=0.8\textwidth]{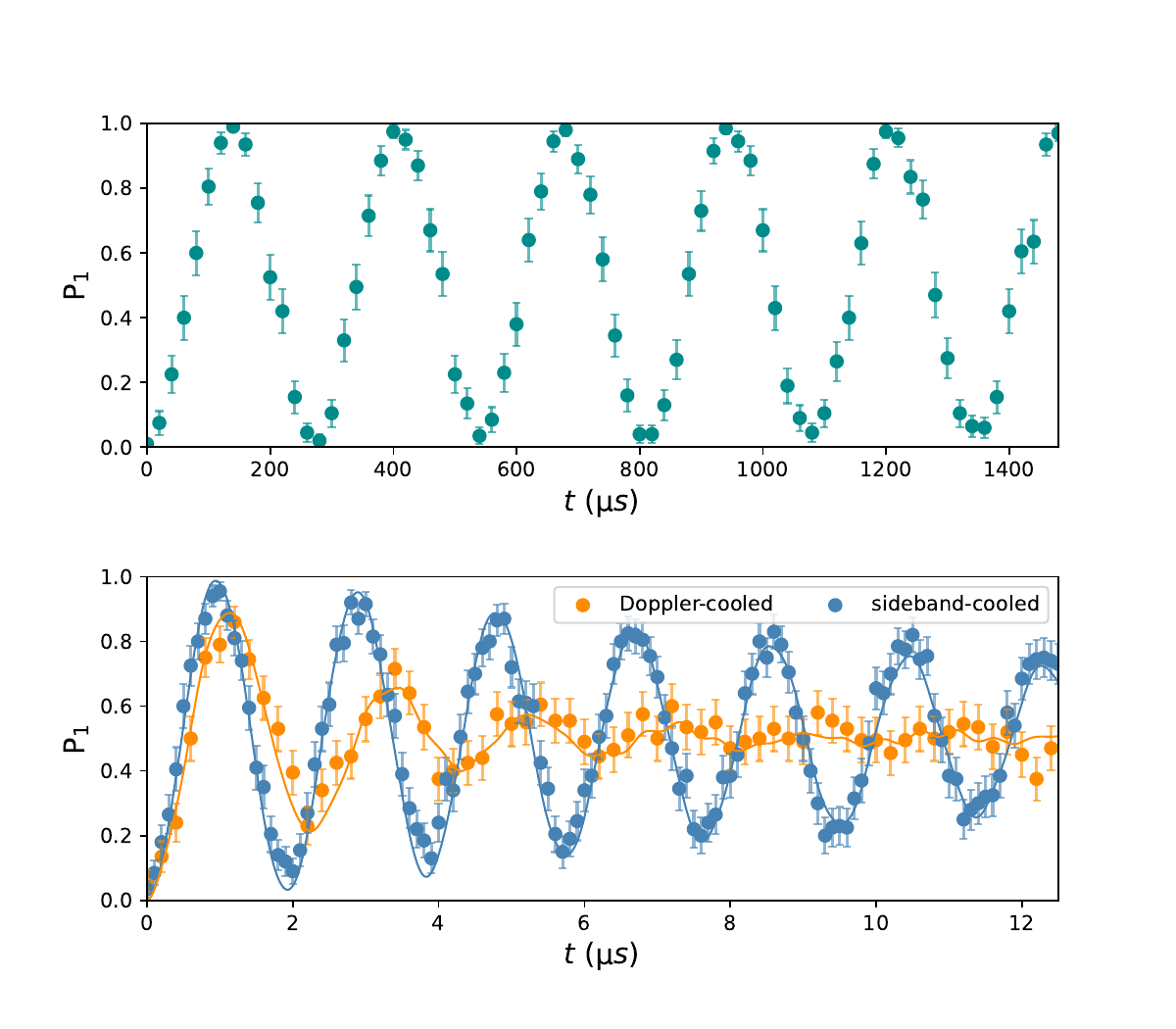}
\caption{Carrier Rabi flopping in different beam configurations. (Top) Co-propagating configuration. This operation is motion-insensitive, flipping the qubit states independent of the vibrational states. (Bottom) Counter-propagating configuration. The qubit state and vibrational states are coupled in this geometry. P$_{1}$ is the population of the state $|1\rangle$. The mean phonon number of the vibrational states are fitted to thermal distributions, $(\bar{n}_{1}, \bar{n}_{2}) = (15.0, 14.0)$ after Doppler cooling, and $(\bar{n}_{1}, \bar{n}_{2}) \approx (4.0, 0.1)$ after sideband cooling (see solid lines). Heating is included in the simulation but has negligible effect during the evolution time.}
\label{fig_rabi}
\end{figure*}

The trap parameters are chosen to set the radial mode secular frequencies of a single ion to $f_{1} = 1.84(1)$ MHz and $f_{2} = 2.11(7)$ MHz, corresponding to $r_1$ and $r_2$ in figure~\ref{fig_experiment_scheme}, respectively. The ion height above the chip surface is 100 µm. Figure~\ref{fig_rabi} exhibits plots of Rabi flopping on the carrier transition driven by Raman beams in different configurations. The top plot was obtained using co-propagating Raman beams, which are insensitive to ion motion. The bottom plot shows Rabi flopping driven by a counter-propagating configuration as shown in figure~\ref{fig_experiment_scheme} (a), where the coupling of Raman beams to the vibrational states is maximal, before and after sideband cooling. 

Due to the limited size of the instruction buffer in our current field programmable gate array (FPGA) controller, we are unable to implement fine-tuned sideband cooling, which is particularly important in systems with high heating rates, to reach the ground state of both modes. Only mode 2 is sufficiently cooled, while mode 1 is cooled intermittently. With a Rabi frequency of 545 kHz, the mean phonon number of each mode before and after sideband cooling are fitted to $(\bar{n}_{1}, \bar{n}_{2}) = (15.0, 14.0)$ and $(4.0, 0.1)$, respectively, assuming thermal distributions (see the solid lines in the bottom plot of figure~\ref{fig_rabi}). We expect to achieve ground state cooling of both modes once we are able to execute more instructions on an upgraded controller.

\begin{figure*}[ht]
\centering
\includegraphics[width=0.5\textwidth]{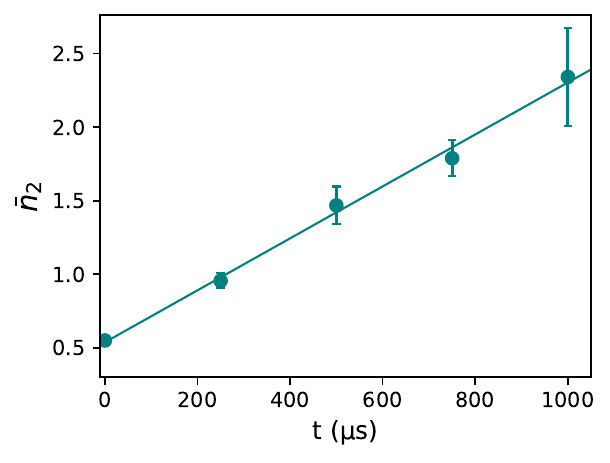}
\caption{Heating rate measurement. The heating rate of mode 2, $\dot{\bar{n}}_2$, is measured using the sideband ratio method. $t$ is the delay time during which heating occurs following the cooling of the ion. The fitted heating rate is $\dot{\bar{n}}_2=1.7(7) \times 10^{3}$ quanta $\cdot \ \mathrm{s}^{-1}$.}
\label{fig_heating}
\end{figure*}

The heating rate of mode 2 is measured to be $\dot{\bar{n}}_2=1.7(7) \times 10^{3}$ quanta $\cdot \ \mathrm{s}^{-1}$ using the sideband ratio method~\cite{Diedrich_1989}. We observed that the dependence of the heating rate on the secular frequency appears random, without an evident scaling pattern of $f^{-\alpha}$ over the range $f_{2}=$ 1.60(9) to 2.11(7) MHz, suggesting that technical noise predominantly influences our system~\cite{Sedlacek_2018}. In addition to reducing technical noise in our system, we are currently investigating other potential sources of heating, including imperfect gold coating on the sidewalls of the slot structure~\cite{Blain_2021}. The partially exposed silicon surfaces on the sidewalls of the slot structure may contribute to the heating rate measured in our system~\cite{Blain_2021}, as space charge regions~\cite{Ghibaudo_1987, Izpura_2007}, generation-recombination processes~\cite{Bonani_1999}, and Schottky barriers~\cite{Hsu_1971} are known sources of electric field noise.

% Note that semiconductor charging is effectively mitigated even with the incomplete protective gold layer shown in figure~\ref{fig_scallop_smoothing} (d). We believe that switching to a more highly conductive silicon wafer (p-type doping concentration of $10^{15} \ \mathrm{cm}^{-3}$ to $10^{19} \ \mathrm{cm}^{-3}$) has also helped by lowering the Fermi level, hence reducing the number of optically excitable defect states at the surface~\cite{Lee_Chung_2024}. 

\subsection{M{\o}lmer-S{\o}rensen gate}
\begin{figure*}[ht]
\centering
\includegraphics[width=1\textwidth]{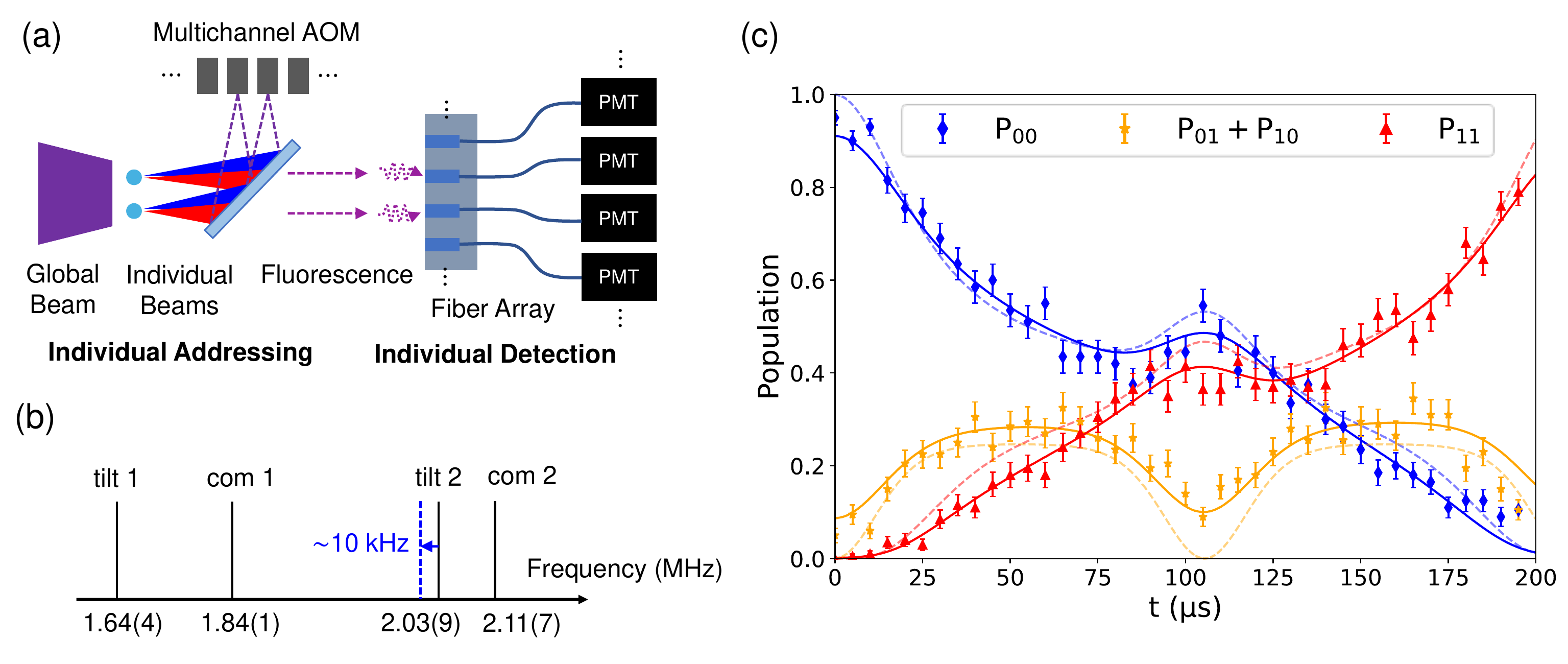}
\caption{Implementation of the M{\o}lmer-S{\o}rensen gate. (a) The experimental set up for individual addressing using a multichannel acousto-optic modulator (AOM), and detection via a fiber array and photomultiplier tubes (PMTs). (b) The blue sideband frequencies of the two-ion chain. The tilt mode of mode $r_2$ is used for the operation due to its low heating rate of 6.(2)$\times 10^{1}$ quanta $\cdot \ \mathrm{s}^{-1}$. The initial mean phonon number is $\bar{n}_\mathrm{tilt,2}=0.2$ assuming a thermal distribution. (c) Result of the M{\o}lmer-S{\o}rensen gate, showing the populations of the two-qubit states. P$_{00}$, P$_{01}$, P$_{10}$, and P$_{11}$ are the populations of the two-qubit states $|00\rangle$, $|01\rangle$, $|10\rangle$, and $|11\rangle$. The solid lines are fits to the data reflecting the limited detection fidelity. The dashed lines are obtained assuming perfect detection fidelity.}
\label{fig_MS}
\end{figure*}

A M{\o}lmer-S{\o}rensen gate is performed with two ions~\cite{Sorensen_1999}, where beams 1 and 2 act as the individual and global beams, respectively. As shown in figure~\ref{fig_MS} (a), the individual beams are controlled by a multichannel acousto-optic modulator (AOM) manufactured by L3Harris Technologies, Inc. The qubit states are individually detected via a fiber array, with each fiber connected to a dedicated Hamamatsu photomultiplier tube (PMT). Due to photon loss in the setup for multi-qubit detection, primarily due to high ultraviolet absorption in the low-OH fibers, the detection fidelity was limited to $\sim 94\%$. This measurement error can be reduced by switching to a high-OH fiber array and optimizing the optical power and detuning of the detection beam.

The heating rates of the center-of-mass (com) and tilt modes of the two-ion chain along $r_2$ (see figure~\ref{fig_experiment_scheme}) are measured to be 2.3(6)$\times 10^{3}$ and 6.2$\times 10^{1}$ quanta $\cdot \ \mathrm{s}^{-1}$, respectively. Therefore, we operate at a frequency slightly red-detuned from the tilt mode, as shown in figure~\ref{fig_MS} (b). After sideband cooling, we implement a M{\o}lmer-S{\o}rensen gate, the result of which is presented in figure~\ref{fig_MS} (c). The data are fitted with Expans-ION, a simulator developed for the exact simulation of operations on an ion chain~\cite{expans-ion}. The fitted Rabi frequencies are 49.9(7) kHz for both ions, with a negative detuning of 9.4(6) kHz with respect to the tilt mode. The initial mean phonon number is fitted as $\bar{n}_\mathrm{tilt,2}=0.2$. The solid and dashed lines correspond to fitted curves with and without consideration of limited detection fidelity, respectively. The measurement errors are accounted for by incorporating $\mathrm{P}(1|0)$ and $\mathrm{P}(0|1)$, which represent the probabilities of misidentifying states $|0\rangle$  as $|1\rangle$ and $|1\rangle$  as $|0\rangle$, respectively. They are fitted as $\mathrm{P}(1|0)= 0.04(5)$ and $\mathrm{P}(0|1)= 0.06(1)$, in agreement with the overall detection fidelity.

\section{Conclusion}
\label{sec_conclusion}
A new fabrication process has been developed to mitigate semiconductor charging, as demonstrated by a chip featuring metallic shielding on all exposed silicon. Silicon-induced stray fields are observed to be significantly reduced in comparison with data from an unshielded chip. Using the shielded chip, we observe stable ion motion, and subsequently conduct motion-sensitive quantum operations, such as sideband cooling and two-qubit gates, that were not feasible in the absence of the gold layer. Additionally, fabrication techniques that produce complex chip outlines reduce laser scattering from the chip surface, further contributing to the suppression of photoinduced charging. The observed heating rate is primarily attributed to technical noise and some unknown sources, such as imperfect shielding of the sidewalls of the slot structure. In the context of ongoing efforts to develop increasingly complex and scalable silicon-based ion-trap chips, the results presented here offer an important means to address a significant source of infidelity inherent to semiconductor materials.

\section{Acknowledgement}
\label{sec_acknowledgement}
This work has been partially supported by the Institute for Information \& communications Technology Planning \& Evaluation (IITP) grant funded by the Korean government (MSIT) (No. 2022-0-01040), the National Research Foundation of Korea (NRF) grant funded by the Korean government (MSIT) (No. 2020R1A2C3005689, No. RS-2024-00442855, No. RS-2024-00413957). Also, this research was funded in part by the Austrian Science Fund (FWF) grants 10.55776/Q4 and 10.55776/W1259. For open access purposes, the author has applied a CC BY public copyright license to any author-accepted manuscript version arising from this submission.

\section*{References}
\bibliographystyle{iopart-num}
\bibliography{reference}

\end{document}